\documentclass[conference]{IEEEtran}
\IEEEoverridecommandlockouts
\usepackage{cite}
\usepackage{algorithmic}
\usepackage{xcolor}
\usepackage{enumerate}
\usepackage{amsfonts,amsmath,amssymb,amsfonts}
\usepackage{amsthm}
\usepackage{algorithmic}
\usepackage{optidef}
\usepackage{algorithm}
\usepackage{mathrsfs}
\usepackage{graphicx}
\usepackage{textcomp}
\usepackage{subfigure}
\usepackage{epstopdf}
\usepackage[squaren]{SIunits}
\usepackage[section]{placeins}%

\def\BibTeX{{\rm B\kern-.05em{\sc i\kern-.025em b}\kern-.08em
    T\kern-.1667em\lower.7ex\hbox{E}\kern-.125emX}}
\begin{document}
\newtheorem{theorem}{\textbf{Theorem}}
\newtheorem{lemma}{\textbf{Lemma}}
\bibliographystyle{IEEEtran}

\title{Transfer Learning Guided Noise Reduction for Automatic Modulation Classification}
\author{\IEEEauthorblockN{Zelin Ji\IEEEauthorrefmark{1},
Shuo Wang\IEEEauthorrefmark{2}, Kuojun Yang\IEEEauthorrefmark{2}, Qinchuan Zhang\IEEEauthorrefmark{2}, Peng Ye\IEEEauthorrefmark{2}}
\IEEEauthorblockA{\IEEEauthorrefmark{1}Shenzhen Institute for Advanced Study, University of Electronic Science and Technology of China, China\\
\IEEEauthorrefmark{2} School of Automation Engineering, University of
 Electronic Science and Technology of China, China\\
Email: zelinji77@gmail.com, 202211060911@std.uestc.edu.cn, \{yangkuojun, zhangqc,
yepeng\}@uestc.edu.cn}
}
\maketitle

\begin{abstract}
Automatic modulation classification (AMC) has emerged as a key technique in cognitive radio networks in sixth-generation (6G) communications. AMC enables effective data transmission without requiring prior knowledge of modulation schemes. However, the low classification accuracy under the condition of low signal-to-noise ratio (SNR) limits the implementation of AMC techniques under the rapidly changing physical channels in 6G and beyond. This paper investigates the AMC technique for the signals with dynamic and varying SNRs, and a deep learning based noise reduction network is proposed to reduce the noise introduced by the wireless channel and the receiving equipment. In particular, a transfer learning guided learning framework (TNR-AMC) is proposed to utilize the scarce annotated modulation signals and improve the classification accuracy for low SNR modulation signals. The numerical results show that the proposed noise reduction network achieves an accuracy improvement of over 20\% in low SNR scenarios, and the TNR-AMC framework can improve the classification accuracy under unstable SNRs.
\end{abstract}

\section{Introduction}
The cognitive radio network (CRN) has become a key technique to enhance transmission efficiency and utilize wireless resources in sixth-generation (6G) and beyond communications. Spectrum sensing is the core idea in the CRN, which enables the effective utilization of the idle spectrum of the primary users (PU)~\cite{9247965}. Automatic modulation classification (AMC) for 6G and beyond communications has emerged as a new approach for spectrum sensing in CRN, which monitors the licensed spectrum of PU and classifies the modulation scheme to improve detection accuracy~\cite{9672088}. 

The vital metrics for AMC are the classification accuracy and the classification latency, i.e., the computational complexity. Conventional likelihood-based AMC methods have long observation latency and rely on prior knowledge~\cite{7728143}, which cannot meet the latency requirement for the real-time classifications. Moreover, the channel state information requirement degrades the classification performance under varying wireless channels. To address the above issues, the feature based AMC, especially the deep learning based AMC (DL-AMC) techniques have become a prevailing approach for AMC regarding classification accuracy and robustness~\cite{8302885}. The signal features include the standard deviations amplitude, phase, frequency, wavelet-based features, statistical features, etc~\cite{10262257}, which can be extracted manually by experts, or automatically by the deep neural networks. Deep learning based algorithms have been widely applied to wireless communication scenarios for semantic extraction, resource optimization, channel estimation, massive access, etc~\cite{9247965,8663966,mfrl2023ji}. In the scope of AMC, the DL-AMC method integrates the features from time-domain and frequency-domain, and matches the extracted features to specific modulation schemes~\cite{9672088}. Peng~\textit{et al.}~\cite{8418751} have leveraged signal constellation diagrams with convolutional neural network (CNN) to extract the features of the signals, and Chen~\textit{et al.}~\cite{9580446} have applied a signal-to-matrix operator and CNN to improve the learning performance in the small-sample learning process. Zhang~\textit{et al.}~\cite{9507514} has combined CNN and recurrent neural network to synthesize the features from multiple dimensions, and the parameter estimation and model pruning techniques were adopted to reduce the complexity of the network. 

On the other hand, the DL-AMC suffers from low classification accuracy under low SNRs and unstable channel state information. Some works have made efforts to enhance the classification accuracy under low SNRs. Zhang~\textit{et al.}~\cite{10663973} proposed a divide-and-conquer domain adapter to divide the data into several sub-domain spaces according to SNRs, and encourage the model to learn the domain internally-invariant feature projections. Ding~\textit{et al.}~\cite{10262257} have combined the semantic attribute information from the data-and-knowledge domain, where the merged features have been proved to improve the classification performance under the low SNR scenarios. Hu~\textit{et al.}~\cite{8891763} have proposed a pre-processing method, to reduce the training cost and improve the robustness of the model under uncertain SNRs. However, these works focus exclusively on enhancing performance under fixed SNRs. In practice, the wireless channel exhibits rapid variability and the signals may suffer from different SNRs. In such scenarios, an effective noise reduction approach should be leveraged to elevate the received SNR of the signals, thereby conferring a further benefit upon the modulation classification results. 

In this paper, we propose a transfer learning guided noise reduction for the AMC scheme (TNR-AMC) to enhance the classification performance under low SNR scenarios. A two-stage transfer learning approach is adopted to utilize the scarce annotated modulation signals. The contributions of this paper that address the above challenges are concluded as follows.

\begin{enumerate}
    \item \textbf{Noise reduction module for AMC}: A deep learning based noise reduction module is proposed to enhance the classification accuracy under low SNR scenarios. The noise reduction module guides the neural networks to focus on the raw data, avoids misleading noise, reconstructs high-precision signals, and benefits classification accuracy.
    \item \textbf{Two-stage transfer learning scheme}: To utilize the scarce annotated modulation signals and raw data effectively, a two-stage transfer learning scheme is proposed. Specifically, the noise reduction model is pre-trained by the period signals, which are easy to acquire and label. In the first transfer learning stage, the pre-trained model is transferred to the modulation signals dataset, and achieves the SNR improvement for the aperiodic modulation signals. In the second transfer learning stage, the noise reduction network is integrated with the classification network, further fine-tuning with the classification results. This approach enables the proposed TNR-AMC framework to learn from the unique features of different modulation schemes, while reducing the effect of varying noise.
\end{enumerate}

The rest of this article is organized as follows. Section~\uppercase\expandafter{\romannumeral2} presents the system model of the signal transmission and the modulation classification. In Section~\uppercase\expandafter{\romannumeral3}, the proposed TNR-AMC framework is introduced in detail. The numerical results and performance improvement are illustrated in Section~\uppercase\expandafter{\romannumeral5}. The conclusion is drawn in Section~\uppercase\expandafter{\romannumeral6}.

\section{System Model}
\label{chp6:system}
In this section, we consider a transmission system with various of modulation schemes. We first formulate the data transmission model, and then introduce the AMC model based on the received signal.

\subsection{Data Transmission Model}
Assuming a data sequence $\boldsymbol{x}$ is transmitted over a physical channel $C(\cdot)$, and the received signal sequence is denoted as 
\begin{equation}
   C(\boldsymbol{x}) = \boldsymbol{y} = [y(0), \dots, y(t), \dots, y(T)]
\end{equation}
where $T$ denotes the length of the sequence of signal samples, $y(t)$ represents the received signal, which can be detailed as
\begin{equation}
   y(t) = h(t) \cdot x(t) + n(t), t = 1, \dots, T,
\end{equation}
where $h(t) = \alpha(t) e^{j\omega(t) + \varphi(t)}$ denotes the channel coefficient, $\alpha(t)$ denotes the channel gain, $\omega(t)$ denotes the frequency shift, and $\varphi(t)$ denotes the phase shift introduced by channel. Gaussian noise $n(t) \sim {\cal CN}\left( {0,\sigma_n^2} \right)$ follows the complex Gaussian distribution. 

To deal with the complex signals, the received signals can be preprocessed into an I/Q sequence as 
\begin{equation}
   y(t) \rightarrow [\Re(y(t)), \Im(y(t))],
\end{equation}
where $\Re(y(t))$ and $\Im(y(t))$ are the real and imaginary parts of the signal, which can be received by the in-phase and quadrature (I/Q) channels, respectively. Hence, the received signal sequence $\boldsymbol{y}$ can be expressed in the form of a matrix as
\begin{equation}
\boldsymbol{y} = \Re(\boldsymbol{y}) + \Im(\boldsymbol{y})
= \left[
\begin{array}{ccc}
\Re(y(0)) & \dots & \Re(y(T)) \\
\Im(y(0))  & \dots & \Im(y(T)) 
\end{array}
\right].
\end{equation}

It is noted that the data of the real and imaginary parts usually follows the identical independent distribution (i.i.d.) for each signal, saving the resource for the normalization operation, which usually requires the whole period of the signals and increases the classification latency. The I/Q signals contain all of the essential information that the signal samples carry~\cite{10262257}, which makes the I/Q raw data suitable to be processed and extracted.

\subsection{Modulation Classification Model}
AMC aims to determine the modulation scheme from $\mathcal{K}$ classes based on the received signal $\boldsymbol{y}$. Denoting $P(\cdot)$ is the classification predictor, the predicted modulation scheme can be expressed as
\begin{equation}
   P(C(\boldsymbol{x})) \rightarrow k, k \in \mathcal{K}.
\end{equation}
Hence, the performance of the classification results is determined by two factors, i.e., the quality of the received signal and the accuracy of the classification predictor. Current works mainly focus on optimizing the predictor $P(\cdot)$ and perform the class prediction based on the received signal $\boldsymbol{y}$. However, the quality of the received signal is unpredictable over the fast varying wireless channels, and the performance of the predictor decrease significantly under low SNR scenarios~\cite{10684251}. A natural thought is to filter the noise of the received signal to improve the SNR, and use the de-noising signal to calibrate the errors introduced by the physical channels, so that increasing the performance under unpredictable wireless environments.

However, we face several challenges in implementing the noise reducing module. Conventionally, the noise can be reduced by the band-pass filters. However, this approach requires a high resolution for the frequency domain, which is not robust for the broadband signal transmission of modern communication systems. The high order of the filter significantly increases the computational complexity. The deep learning based approaches can be leveraged to enhance the robustness of the denoising module~\cite{9440746, Liu2020ModulationRW}. Another challenge is caused by the lack of real-world modulation signals over fast-varying channels, where the network can hardly learn from the aperiodic modulation signals. Additionally, current works only optimize the denoising module and the classification module separately, which may affect the classification accuracy performance. To overcome the above challenges, we propose a transfer learning guided noise reduction module, and combine the noise reduction module with the classification module to enhance the overall performance of the system.


\section{The Transfer Learning Guided Noise Reduction for AMC Framework}
\label{algorithm}
\begin{figure}[t]
\centering
\includegraphics[width=\columnwidth]{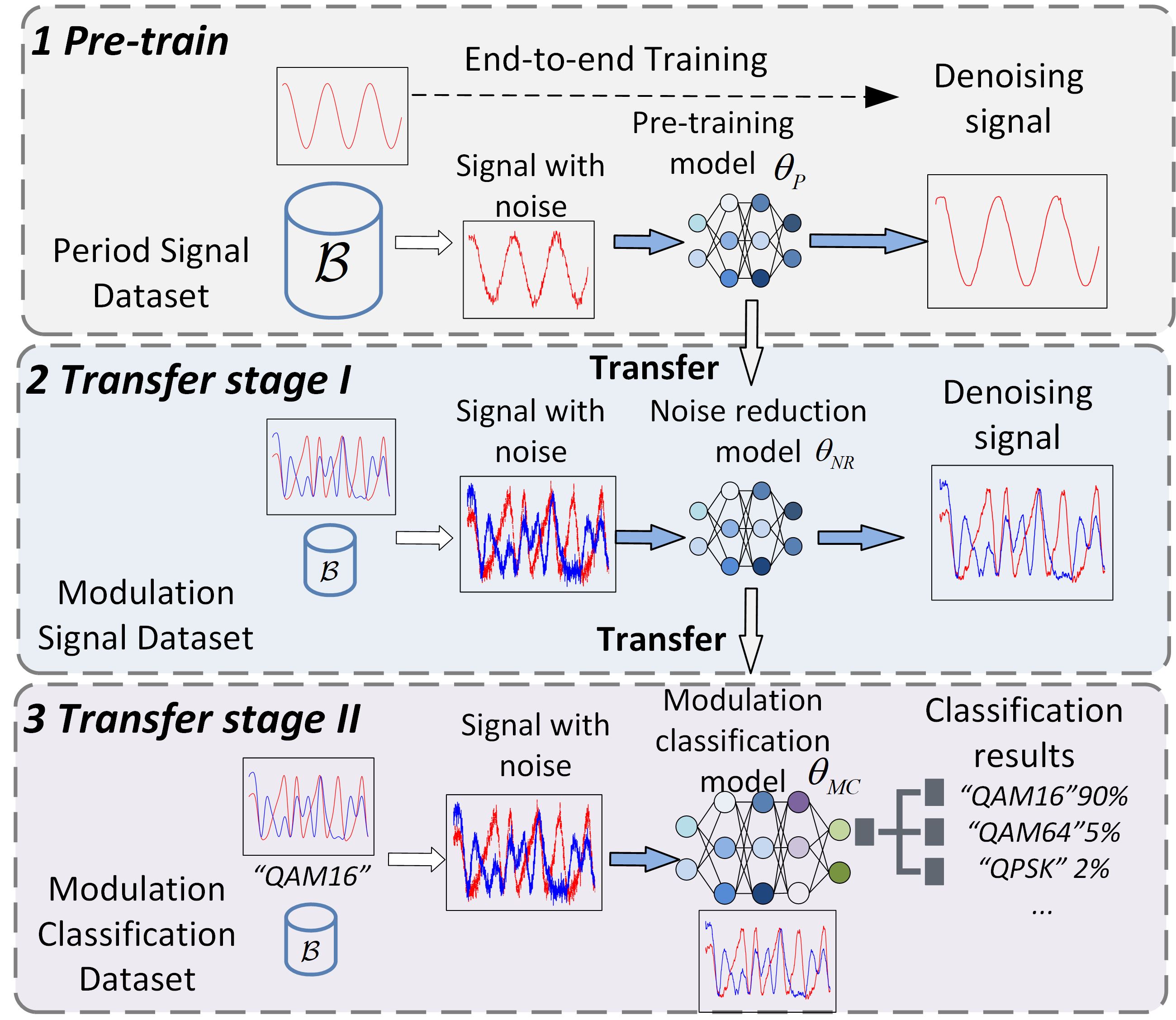}
\caption{The proposed anomaly detected procedure.}
\label{procedure}
\end{figure}

In this section, the implementation of the proposed transfer learning guided noise reduction for the AMC framework (TNR-AMC) is presented. The framework includes several procedures, including pre-training of the noise reduction for the period signals, transfer learning of the noise reduction model for the modulation signals, the transfer learning for the modulation classification. The whole process is illustrated in Fig.~\ref{procedure}. In the following paragraphs, the details of each module will be introduced.

\subsection{Transfer Learning Guided Noise Reduction Module}

The deep learning based denoising approaches provide a promising solution for intelligent noise reduction for arbitrary signals. However, enormous labeled data is required to train a robust and effective denoising model due to the low data efficiency issue of deep learning. Specifically, for the aperiodic modulation signals, the denoising feature of the original data is harder to be extracted than periodic data. Meanwhile, the scarcity of annotated modulation data worsens the situation, further restricting the application of the deep learning based denoising methods for the communication signals. 

An effective solution to the limited dataset is transfer learning, which can transfer knowledge from different but related task domains, so that reducing the dependence on enormous target-domain data and improving the performance of the objective tasks~\cite{9134370}. In the proposed scenario, the noise reduction for practical communication signals with fast varying channels can be an analogy to the denoising for period signals, which is much easier to acquire and label. 

The key idea of noise reducing is to minimize the reconstruction error of the neural network with the original signal. To deal with the 1-dimensional time-domain data stream, we apply the gated recurrent unit (GRU) network to extract the temporal information. The output of the GRU network can be denoted as $\hat{\boldsymbol{x}} = [\hat{x}(0), \dots, \hat{x}(T) ]$, and the reconstruction error can be measured by the mean square error (MSE) as
\begin{equation}
   {\cal L}_{P} = \frac{1}{T}\sum^T_{t=0}\left\Vert \hat{x}(t) - x(t) \right\Vert^ 2.
\label{loss_P}
\end{equation}

To overcome the shortage of labeled modulation data, the denoising ability of the pre-trained model is extended to the modulation data in the first transfer learning stage. The I/Q signals follow the i.i.d. and can be processed simultaneously, and the MSE loss for the modulation signal reconstruction can be denoted as

\begin{align}
       {\cal L}_{NR} = \frac{1}{T}\sum\limits^T_{t=0}\left\{\left\Vert \Re(\hat{x}(t)) - \Re(x(t)) \right\Vert^2 \right. \notag \\
  \left. + \left\Vert \Im(\hat{x}(t)) - \Im(x(t)) \right\Vert^ 2 \right\},
\label{loss_NR}
\end{align}
where $\Re(\boldsymbol{\hat{x}}) = [\Re(\hat{x}(0)), \dots, \Re(\hat{x}(T)) ]$ and $\Im(\boldsymbol{\hat{x})} = [\Im(\hat{x}(0)), \dots, \Im(\hat{x}(T)) ]$ represent the denoising signal of $\Re(\boldsymbol{y})$ and $\Im(\boldsymbol{y})$, respectively.

\begin{algorithm}[t]
\caption{Transfer learning guided noise reduction framework.}
\label{algo1}
\begin{algorithmic}[1]
\STATE \textbf{Input}: The period signal dataset ${\cal B}_P$, the modulation denoising dataset ${\cal B}_{NR}$, the modulation classification dataset ${\cal B}_{MC}$, I/Q sequence;\
\STATE Initialize the noise reduction network $\theta_{P}$, $\theta_{NR}$ and classification network $\theta_{MC}$;\
\STATE \text{\textbf{Pre-train}}:\
\FOR{each training epoch}
\STATE Input the received signal $\boldsymbol{y}$ with Gaussian noise to the GRU network $\theta_P$ and acquire the output $\hat{\boldsymbol{x}}$;\
\STATE Calculate the reconstruction error ${\cal L}_{P}$ by~(\ref{loss_P});\
\STATE Train the neural network by $\theta_P \leftarrow \theta_P + \alpha_P\nabla_{\theta_P}{\cal L}_{P}$ ;\
\ENDFOR
\STATE \text{\textbf{Transfer stage \uppercase\expandafter{\romannumeral1}}}:\
\FOR{each transfer learning epoch}
\STATE Transfer the pre-training network $\theta_P$ to noise reducing network $\theta_{NR}$ for modulation data;\
\STATE Input the modulation signal $\boldsymbol{y}$ with Gaussian noise to the GRU network $\theta_{NR}$ and acquire the output $\hat{\boldsymbol{x}}$;\
\STATE Calculate the reconstruction MSE loss ${\cal L}_{NR}$ by~(\ref{loss_NR}) and train the neural network by $\theta_{NR} \leftarrow \theta_{NR} + \alpha_{NR}\nabla_{\theta_{NR}}{\cal L}_{NR}$;\
\ENDFOR
\STATE \text{\textbf{Transfer stage \uppercase\expandafter{\romannumeral2}}}:\
\FOR{fine-tuning epoch}
\STATE Transfer the noise reduction network $\theta_{NR}$ to noise reducing network $\theta_{MC}$ as the first several layers;\
\STATE Input the modulation signal $\boldsymbol{y}$ with Gaussian noise to the classification network $\theta_{MC}$ and acquire the output $P(k)$;\
\STATE Calculate the cross-entropy and MSE loss ${\cal L}_{MC}$ by~(\ref{loss_MC}) and fine-tune the neural network by $\theta_{MC} \leftarrow \theta_{MC} + \alpha_{MC}\nabla_{\theta_{MC}}{\cal L}_{MC}$;\
\ENDFOR
\STATE \textbf{Return}: TNR-AMC model $\theta_{MC}$.\
\end{algorithmic}
\end{algorithm}

\begin{figure*}[t]
\centering
\includegraphics[width=1.8\columnwidth]{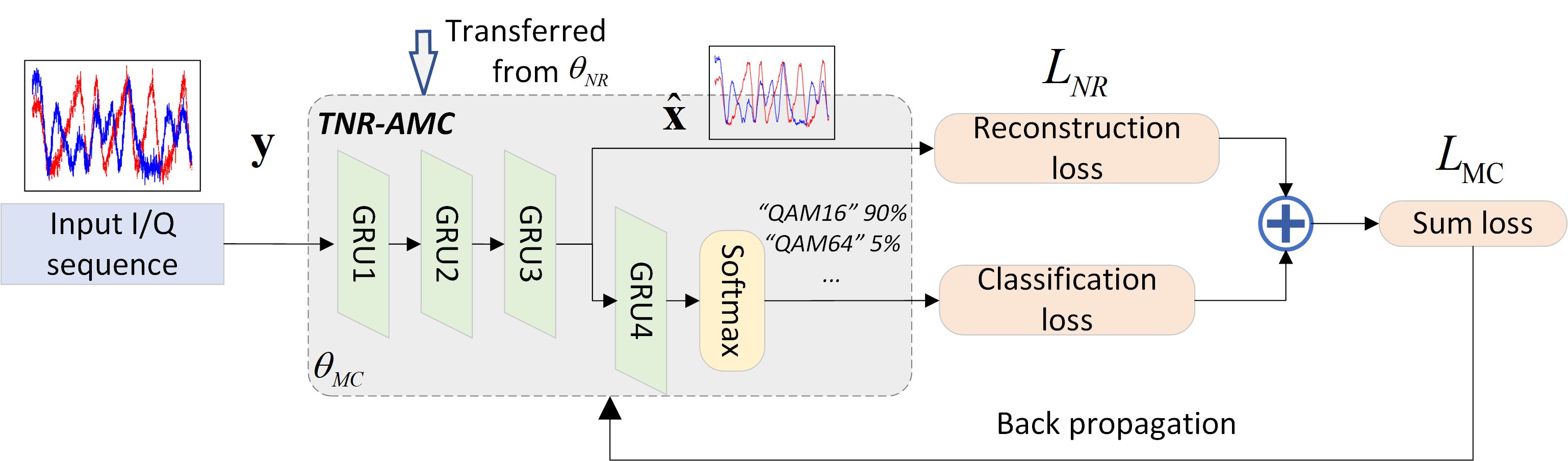}
\caption{The proposed network structure of the TNR-AMC.}
\label{structure}
\end{figure*}

\subsection{Transfer Learning for Modulation Classification}
After transferring the noise reduction model to the modulation signals, the SNR of the signals can be improved significantly and classified accurately. However, the separate design of the noise reduction model and the classification model may degrade the overall performance of the classification accuracy. To address this issue, the second stage of the transfer learning is designed to merge the noise reduction network and the classification network, and fine-tune the noise reduction network and the classification network simultaneously. The detailed structure of the TNR-AMC framework is illustrated in Fig.~\ref{structure}.

The loss function of the TNR-AMC network includes two parts, the reconstruction loss and the classification loss. The reconstruction loss takes the network output of the first two GRU layers which are transferred from $\theta_{NR}$, and calculates the MSE with the original signal $\boldsymbol{x}$. The classification loss is calculated by the cross-entropy of the softmax probability. The sum loss of the TNR-AMC network can be expressed by

\begin{align}
   {\cal L}_{\rm{MC}} = &w_{NR}\frac{1}{T}\sum\limits^T_{t=0}\left\{\left\Vert \Re(\hat{x}(t)) - \Re(x(t)) \right\Vert^2 \notag \right.\\
   & \left. + \left\Vert \Im(\hat{x}(t)) - \Im(x(t)) \right\Vert^ 2 \right\} \notag\\ 
   &-w_{MC}\frac{1}{K}\sum^K_{k=1}\rho_{k}\log(P(k)),
\label{loss_MC}
\end{align}
where $w_{NR}$ and $w_{MC}$ represent the weights for the noise reduction and modulation classification, respectively, $K$ represents the number of anomaly categories, $P(k)$ denotes the output softmax probability of the signal belongs to the category $k$, $\rho_{k} = 1$ if the real category of signal belongs to $k$, otherwise $\rho_{k} = 0$. 

Finally, the loss is back propagated to train the TNR-AMC network, which is fine-tuned and acquires the ability to denoise and classify the signals with low and unstable SNRs. The algorithm is detailed in \textbf{Algorithm~\ref{algo1}}.

\section{Numerical Results}
\label{results}

\begin{figure} 
\centering 
    \includegraphics[width=\columnwidth]{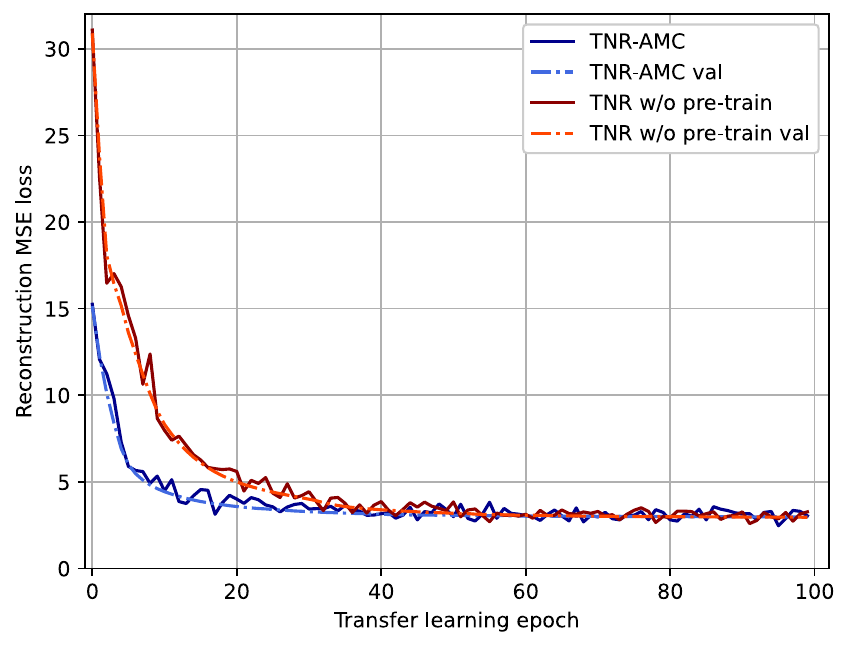}
  \caption{Training loss comparison.}
  \label{transfer_loss}
\end{figure}

\begin{figure} 
\centering 
\subfigure[Training loss.]{
    \includegraphics[width=\columnwidth]{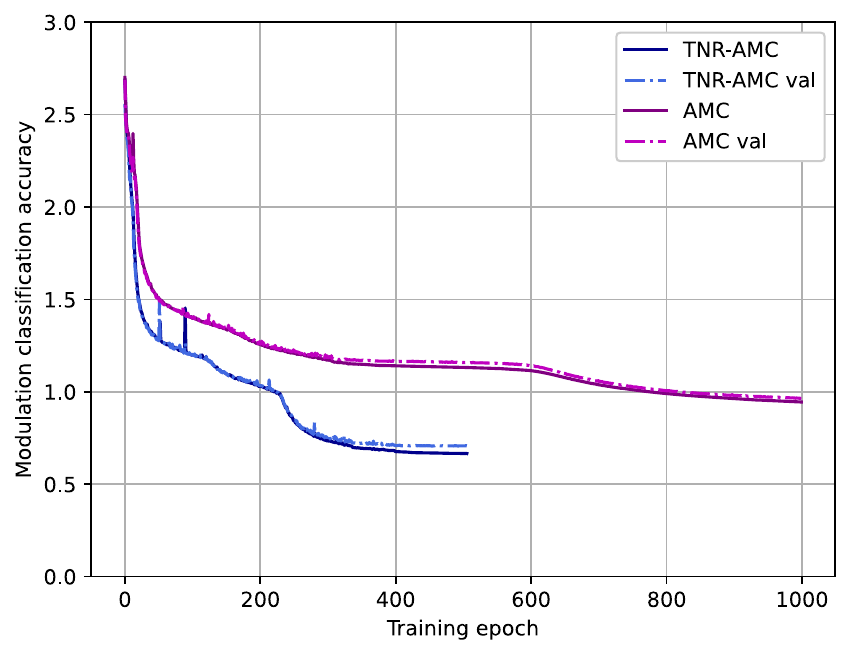}
    \label{MC_loss} }
\subfigure[Training accuracy.]{
    \includegraphics[width=\columnwidth]{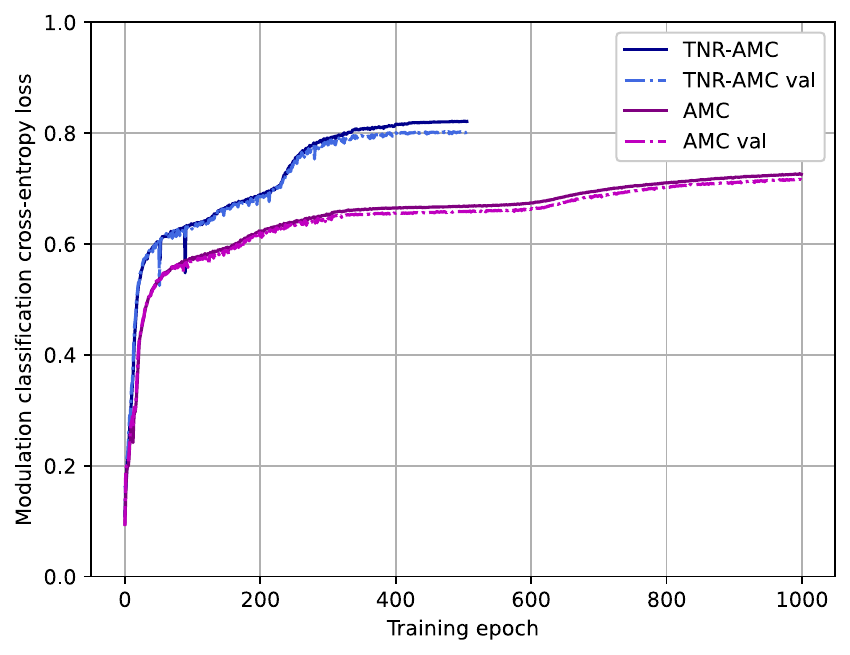}
    \label{MC_acc} 
}
  \caption{The training loss and the accuracy comparison of the proposed TNR-AMC framework and conventional AMC scheme. The input signals are with the noise from -10 to +18 dB.}
  \label{AMC_comparison}
\end{figure}

In this section, the performance of the proposed TNR-AMC framework is demonstrated by simulation results. We first train the noise reduction model using the sine, square, triangle, and sawtooth signals. The signals comprise 5-10 periods with the random SNR from -10 to +18 dB. Note that the SNR not only varies between different signals but also varies within each signal, reflecting the rapid fluctuations in channel states. The dataset contains the signal with 10,000 signal samples of 1280 length for each signal. Then, the noise reduction model is transferred to the modulation signal dataset. 

Different from other works, to acquire the original transmit signal $\boldsymbol{x}$, the publicly available modulation signal dataset RadioML2016.10a~\cite{2016a} cannot be leveraged due to the lack of the perfect signal without the distortion by the dynamic channels. Hence, we build up a dataset using the digital modulation source SMW200A from Rohde \& Schwarz. The channel is modeled by the additive white Gaussian noise (AWGN) channel, with the SNR in the -10 to +18 dB range. The modulation schemes include BPSK, QPSK, 8PSK, 16QAM, and 64QAM, each with 1000 signal samples of 1280 length.


It is worth noting that the proposed TNR-AMC framework also works well with other deep neural networks, e.g., long short-term memory (LSTM), RNN, CNN, residual network (ResNet), etc. In this paper, we apply GRU to implement the TNR-AMC framework because it can achieve good performance with low computational complexity, and is easy to train. To further demonstrate the effectiveness of the proposed TNR-AMC framework, we also compare the testing accuracy performance on RadioML2016.10a with the following benchmarks.
\begin{itemize}
\item {\textbf{NR w/o pre-train}: The network is only trained with the modulation data for noise reduction and classification, without the pre-train process with the period signal dataset.}
\item {\textbf{AMC w/o transfer}: The network is only trained with the modulation data for classification direct, which is the most common way for the AMC techniques~\cite{ZHANG2022103650}.}
\end{itemize}

The GRU network $\theta_{P}$ and $\theta_{NR}$ for the noise reduction contains 3 layers, each with 32 neurons. The first 3 layers of the modulation classification network $\theta_{MC}$ have an identical structure with the noise reduction model, accommodating the transferred model $\theta_{NR}$. The learning rate $\alpha_P$ and $\alpha_{NR}$ are set to $\rm{e}^{-3}$, and $\alpha_{MC}$ starts from $\rm{e}^{-3}$, and decreases until $\rm{e}^{-5}$ with the fine-tune process. The weights $w_{NR}$ and $w_{MC}$ are set to 0.1 and 0.9, respectively.

We first investigate the performance of the noise reduction module. Fig.~\ref{transfer_loss} shows the training and validation loss of the proposed TNR-AMC framework and the conventional NR algorithm without the pre-training stage. It can be observed that the proposed transfer learning based TNR scheme achieves much lower MSE loss at the beginning of the training stage. Meanwhile, the TNR-AMC scheme converges in only 20 epochs while the TNR scheme without transfer learning requires more than 40 epochs to converge.

\begin{figure} 
\centering 
    \includegraphics[width=\columnwidth]{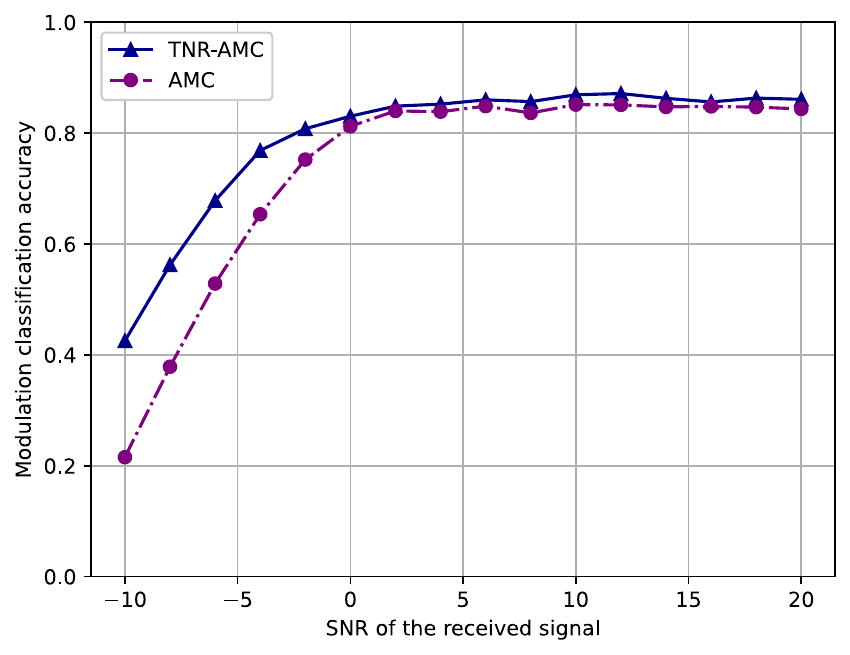}
  \caption{Classification accuracy over SNRs.}
  \label{MC_SNR}
\end{figure}

Subsequently, we observe the performance of the fine-tuning stage when transferring the noise reduction model to the classification model in Fig.~\ref{AMC_comparison}. Although TNR-AMC and AMC have high training loss and low classification accuracy at the beginning of the training, the TNR-AMC algorithm achieves much lower training loss and higher accuracy with more training epochs. Moreover, it is clear that the TNR-AMC scheme also achieves early convergence at around 500 epochs with over 80\% classification accuracy, while the AMC scheme converges at 1000 epochs with 71\% classification accuracy. Fig.~\ref{MC_SNR} explains the improvement of the accuracy. Although the performance of TNR-AMC does not differ from the AMC significantly, the gap between these two schemes is large from -10 to 0 dB, highlighting the effectiveness of the noise reduction model. The promising results lay a foundation for applying the TNR-AMC framework in fast-varying channels, especially under scenarios with low SNRs. Compared to the AMC scheme, the TNR-AMC scheme improves the classification by about 20\% at -10 dB and -8 dB, which equals improving the signal's SNR over 3 dB.

\begin{figure} 
\centering 
\subfigure[AMC (acc = 40.9\%)]{
    \includegraphics[width=\columnwidth]{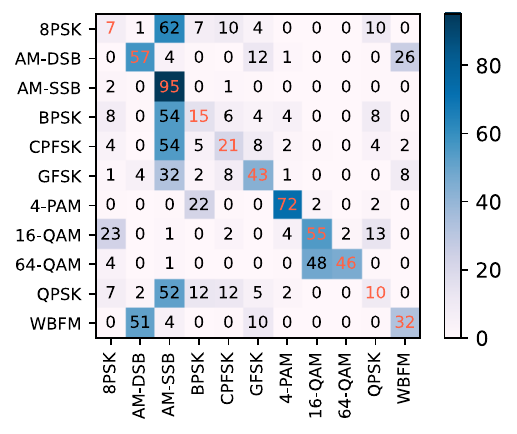}
    \label{confusion}}
\subfigure[TNR-AMC (acc = 56.2\%)]{
    \includegraphics[width=\columnwidth]{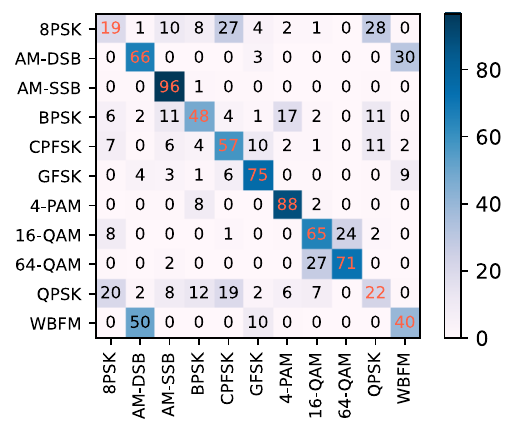}
    \label{denoising_confusion} 
}
  \caption{The confusion matrix comparison of the GRU scheme and the proposed TNR-AMC framework. The SNR of the input signal is -8 dB.}
  \label{confusion_matrix}
\end{figure}

Finally, we verify the effectiveness of the proposed framework using the publicly available modulation signal dataset RadioML2016.10a, where the data with SNR = -8dB is extracted to test the classification accuracy under bad channel situations. Fig.~\ref{confusion_matrix} shows the confusion matrix of the AMC scheme and the proposed TNR-AMC scheme. The accuracy of the proposed TNR-AMC scheme is significantly improved, increasing from 40.9\% to 56.2\%, which verifies the accuracy improvement performance for classifying low SNR signals.

\section{Conclusion}
\label{chp6:conclusion}
In this paper, a transfer learning guided noise reduction framework for automatic modulation classification is proposed.  To achieve effective noise reduction with scarce labelled modulation data, a transfer learning approach is used to transfer the experience of the periodic signal denoising model to aperiodic modulation signals. The noise reduction model is then transferred to the modulation classification network to improve the classification accuracy for signals with low SNRs. Numerical results demonstrated that the proposed transfer learning guided noise reduction for automatic modulation classification framework achieved over 15-25\% accuracy improvement under low SNR scenarios, and the transfer learning based framework is a promising and general solution for the challenging classification tasks in which the features are submerged by the interference.

\bibliography{Reference}
\clearpage
\end{document}